\documentclass[fleqn,usenatbib]{mnras}

\usepackage{newtxtext,newtxmath}
\usepackage[T1]{fontenc}

\DeclareRobustCommand{\VAN}[3]{#2}
\let\VANthebibliography\thebibliography
\def\thebibliography{\DeclareRobustCommand{\VAN}[3]{##3}\VANthebibliography}

\usepackage{graphicx}	% Including figure files
\usepackage{amsmath}	% Advanced maths commands
\title[SB2 FEROS]{Test of binary spectral model on FEROS spectra. First estimation of the fundamental parameters for HD~20784.}

\author[M. Kovalev et al.]{
Mikhail Kovalev,$^{1,2,3,4}$\thanks{E-mail: mikhail.kovalev@ynao.ac.cn}
Sarah Gebruers,$^{5,4}$
Ilya Straumit$^{5,6}$
\\
$^{1}$Yunnan Observatories, China Academy of Sciences, Kunming 650216, China\\
$^{2}$Key Laboratory for the Structure and Evolution of Celestial Objects, Chinese Academy of Sciences, Kunming 650011, China\\
$^{3}$Sternberg Astronomical Institute of M. V. Lomonosov Moscow State University, Leninskie Gory, Moscow 119991, Russia\\
$^{4}$Max Planck Institute for Astronomy, D-69117 Heidelberg, Germany\\
$^{5}$Institute for Astronomy, KU Leuven, Celestijnenlaan 200D bus 2401, 3001 Leuven, Belgium\\
$^{6}$Department of Astronomy, The Ohio State University, Columbus, OH 43210, USA\\
}

\date{Accepted XXX. Received YYY; in original form ZZZ}

\pubyear{2022}

\def\kms{\,{\rm km}\,{\rm s}^{-1}}

\def\feh{\hbox{[M/H]}}

\newcommand{\teff}{T_{\rm eff}}
\newcommand{\rv}{{\rm RV}}
\def\Vmic{V_{\rm mic}}

\def\vsini{V \sin{i}}%\def\vsini{V{\rm \sin }i}
\def\logg{\log{\rm (g)}}
\def\snr{\hbox{S/N}}

\begin{document}
\label{firstpage}
\pagerange{\pageref{firstpage}--\pageref{lastpage}}
\maketitle

\begin{abstract}
We analyse spectra of 26 early-type stars, with suspected binarity, using a binary spectral model adapted for high-resolution FEROS spectra. We confirm seven SB2 candidates (AE Pic, $\epsilon$ Vol, HD 20784, HD 208433, HD 43519, HD 56024, CD-73~375A) and derive mass ratios and spectroscopic parameters for them. We find good agreement with theoretical models. For slightly evolved system HD~20784 we made the first estimation of the fundamental parameters and age $\log{t}=8.5$ yr.  
% This is a simple template for authors to write new MNRAS papers.
% The abstract should briefly describe the aims, methods, and main results of the paper.
% It should be a single paragraph not more than 250 words (200 words for Letters).
% No references should appear in the abstract.
\end{abstract}


%%%%%%%%%%%%%%%%%%%%%%%%%%%%%%%%%%%%%%%%%%%%%%%%%%

%%%%%%%%%%%%%%%%% BODY OF PAPER %%%%%%%%%%%%%%%%%%

\section{Introduction}
Spectral modelling of binary stars is much more complex than for single stars. Generally one needs to model two sets of spectral parameters together with the Doppler shift for each component. Another problem is a correct scaling of the light contribution of each component in the spectrum. Recently \citet{bardy2018} and \cite{m11} developed efficient methods to fit composite spectra with medium resolution ($\lambda/d\lambda\sim 20\,000$) for infrared spectra of the APOGEE \citep{dr14} and Gaia-ESO \citep{Gilmore2012} surveys. \cite{tyc} also developed and tested a binary spectral model for relatively cool stars ($\teff<9000$ K) using visual medium-resolution ($\lambda/d\lambda\sim 7500$) LAMOST-MRS spectra \citep{lamostmrs}. 
\par
In this paper we use a sample of high-resolution spectra of bright B-type stars with suspected binarity to test the binary model for hot stars.
The paper is organised as follows: in Section~\ref{Observations} we describe the observations and methods. Section~\ref{results} presents and discuss our results. In Section~\ref{concl} we summarise the paper and draw conclusions.

\section{Observations \& Methods }
\label{Observations}
\subsection{Observations}

The targets studied here are a subset of the samples from \citet{Pedersen2019} and \citet{Bowman2019}. These two samples contain OB-type pulsators with a diversity of variability detected in the first three sectors of their TESS light curves that were available at that time. It concerns stars with pulsational, binary, rotational, or stochastic low frequency variability, both Galactic and in the LMC, that are therefore interesting for asteroseismic modelling.
For the 166 brightest targets in these samples (V-band magnitude < 11.5\,mag) spectra have been obtained with the FEROS spectrograph \citep{Kaufer1999}. This is the spectrograph attached to the ESO/MPG 2.2m telescope at the La Silla observatory in Chile. It has a resolution of $\sim\,$ 48\,000 and covers a wavelength range of 3600-9200 \AA \ over 39 spectral orders. Each target has been observed at least twice, with one epoch in December 2019 and the other one in February 2020, except for seven targets that were not visible during the second run. The spectra were reduced with the CERES pipeline \citep{Brahm2017} which is a Python package with automated reduction pipelines for 13 different \'echelle spectrographs including FEROS. \citet{gebruers22} describe four adaptations made to the CERES pipeline to improve the reduction process of FEROS spectra resulting in smoother order stitching and the removal of cosmic hits. These authors also performed a least-squares deconvolution analysis \citep[LSD;]{Donati1997} following the method of \citet{Tkachenko2013} to identify binary systems and line profile variability in the spectra of the 166 targets. In total 26 targets were found to be SB2 systems. We list information for these stars in Table~\ref{tab:inp}. A more detailed overview of the sample selection, spectrum reduction with the CERES pipeline and LSD analysis can be found in Sects. 2, 3 and 4.1 of \citet{gebruers22}.  

\begin{table*}
    \centering
    \begin{tabular}{lccccccc}
    \hline
    SIMBAD id&type&R.A.(J2000)&Dec (J2000)&B, mag&V, mag&sp type&N$_{\rm spectra}$\\
    \hline
  HD 225119 & Ro* & 00 03 37.10 & -28 25 01.7 & 8.07 & 8.18 & ApSi & 3\\
  HD 20784 & Ro* & 03 18 24.72 & -55 49 21.1  & 8.26 & 8.28 & B9.5V & 3\\
  HD 29994 & Ro* & 04 38 13.20  & -68 48 42.0 & 8.09 & 8.11 & B8/9V & 3\\
  V* AN Dor & EB* & 04 52 28.24  & -55 41 49.4  & 7.47 & 7.69 & B2/3V & 3\\
  HD 32034 & s*b & 04 55 11.09 & -67 10 10.4  & 9.756 & 9.715 & B9Iae & 2\\
  HD 33599 & Be* & 05 07 12.95 & -61 48 18.3 & 9.56 & 9.37 & B2Vpe & 2\\
  HD 35342B & s*b & 05 18 11.91 & -69 13 03.4 & 11.489 & 11.281 & B0.5I & 2\\
  HD 269606 & s*b & 05 28 49.18 & -66 59 36.3  & 11.382 & 11.43 & B7Ia & 2\\
  HD 43519 & ** & 06 12 32.20  & -61 28 26.6 & 6.58 & 6.64 & B9.5V & 2\\
  HD 45796 & Pu* & 06 24 55.80 & -63 49 41.3 & 6.124 & 6.249 & B6V & 3\\
  V* AE Pic & EB* & 06 31 10.64 & -61 52 46.4 & 5.99 & 6.144 & B2V & 3\\
  HD 47620 & * & 06 35 56.94 & -58 08 55.1  & 9.94 & 10.0 & B6V & 2\\
  HD 48559 & * & 06 40 21.88 & -59 07 23.5  & 7.68 & 7.78 & B6III & 3\\
  HD 53327 & * & 07 00 43.19 & -61 08 30.5 & 7.76 & 7.72 & B9.5V & 3\\
  HD 53921 & ** & 07 03 15.10 & -59 10 41.2 &  &  & B9III+B8V & 3\\
  HD 54967 & Pu* & 07 07 16.28 & -59 42 59.1 & 6.35 & 6.47 & B4V & 3\\
  %HD 55478 & Ro* & 07 07 46.9667342448 & -67 56 12.706546668 & 7.95 & 8.06 & B8III & 2\\
  HD 56024 & * & 07 11 50.19  & -58 51 02.4 & 8.71 & 8.71 & B9V & 2\\
  CD-73 375A & Ro* & 07 35 21.66 & -74 16 29.8  & 7.04 & 7.08 & B9IV & 2\\
  HD 61644 & EB* & 07 36 25.25  & -61 52 25.9 &  &  & B5/6IV & 3\\
  HD 62688 & ** & 07 39 31.60  & -70 26 44.1 & 8.99 & 9.05 & B8V & 2\\
  HD 63204 & a2* & 07 43 36.96 & -63 23 57.7 & 8.27 & 8.31 & ApSi & 3\\
  V* V356 Car & EB* & 07 58 02.91  & -60 36 53.2  & 7.55 & 7.59 & B9.5IVpSi & 3\\
  * $\epsilon$ Vol & SB* & 08 07 55.79  & -68 37 01.4  & 4.27 & 4.398 & B5III & 3\\
  * $\kappa$01 Vol & * & 08 19 48.96  & -71 30 53.7 & 5.245 & 5.318 & B9III/IV & 4\\
  HD 198174 & Ro* & 20 49 17.61  & -25 46 52.5 & 5.766 & 5.849 & B8II & 1\\
  HD 208433 & ** & 21 56 55.55 & -35 21 35.4 & 7.43 & 7.44 & B9.5V & 2\\
    \hline
    
    \end{tabular}
    \caption{Observed sample information extracted from SIMBAD sorted by R.A. Types are: *-star, ** -double or multiple star, a2* - $\alpha$2 CVn, Ro* - rotation variable, Be* -B emission star, s*b -blue subgiant, Pu* -pulsating variable, EB* -eclipsing binary, SB* -spectral binary.}
    \label{tab:inp}
\end{table*}

\subsection{Spectral fitting}
\label{sec:maths} % used for referring to this section from elsewhere

Our spectroscopic analysis includes two consecutive stages: 
\begin{enumerate}
    \item analysis of individual observations by binary and single-star spectral models, where we normalise the spectra and make a rough estimation of the spectral parameters, see brief description in Section~\ref{sec:ind}. This is a further development of the method presented in \citet{m11}. 
    \item simultaneous fitting of multi-epochs with a binary spectral model, using constraints from binary dynamics and values from the previous stage as an input \citep{tyc}, see Section~\ref{sec:multi}.
\end{enumerate}

\subsubsection{Individual spectra}
\label{sec:ind}

We use the spectral model from \citet{gebruers22} as a single-star model.
This is the generative neural network model $the~Payne$ \citep{ting2019}, trained on a grid of synthetic spectra generated using the GSSP (Grid Search in Stellar Parameters) software \citep{2015A&A...581A.129T}. GSSP uses the SynthV code for spectral synthesis \citep{1996ASPC..108..198T}, which is based on the LTE (local thermodynamic equilibrium) approach and uses a list of spectral lines from the VALD database \citep{2015PhyS...90e4005R}. For more details please check \citet{gebruers22}.
\par
The normalised binary model spectrum is generated as a sum of two Doppler-shifted normalised single-star model spectra ${f}_{\lambda,i}$ scaled according to the difference in luminosity, which is a function of the $\teff$ and stellar size. We assume both components to be spherical and use following equation:    

\begin{align}
    {f}_{\lambda,{\rm binary}}=\frac{{f}_{\lambda,2} + k_\lambda {f}_{\lambda,1}}{1+k_\lambda},~
    k_\lambda= \frac{B_\lambda(\teff{_{,1}})~M_1}{B_\lambda(\teff{_{,2}})~M_2} 10^{\logg_2-\logg_1}
	\label{eq:bolzmann}
\end{align}
 where  $k_\lambda$ is the luminosity ratio per wavelength unit, $B_\lambda$ is the black-body radiation  (Plank function), $\teff$ is the effective temperature, $\logg$ is the surface gravity and $M$ is the mass. Throughout the paper we always assume the primary star to be brighter.%more massive:  $q \geq 1$.of each component , defined separately for blue and red arms of the spectrum
\par
The binary model spectrum is later multiplied by the normalisation function, which is a linear combination of the first ten Chebyshev polynomials \citep[similar to][]{kovalev19}. The resulting spectrum is compared with the observed one using the \texttt{scipy.optimise.curve\_fit} function, which provides optimal spectral parameters, radial velocities (RV) of each component plus mass ratio $q={M_1}/{M_2}$ and ten coefficients of the Chebyshev polynomials. We keep metallicity equal for both components. In total we have 23 free parameters for a binary fit. We estimate goodness of the fit parameter by reduced $\chi^2$:

\begin{flalign}
\label{eq:chi2}
   \chi^2 =\frac{1}{N-p} \sum \left[ \left({f}_{\lambda,{\rm observed}}-{f}_{\lambda,{\rm model}}\right)/{\sigma}_{\lambda}\right]^2
\end{flalign}
where $N$ is a number of wavelength points in the observed spectrum and $p$ is a number of free parameters. To explore the whole parameter space and to avoid local minima we run the optimisation six times with different initial parameters of the optimiser. We select the solution with minimal  $\chi^2$ as a final result. 
\par 
Additionally, every spectrum is analysed by a single star model, which is identical to a binary model when the parameters of both components are equal, so we fit only for 16 free parameters. Using this single star solution, we compute the difference in reduced $\chi^2$ between two solutions and the improvement factor, computed using Equation~\ref{eqn:f_imp} similar to \cite{bardy2018}. This improvement factor estimates the absolute value difference between two fits and weights it by the difference between two solutions.

\begin{align}
\label{eqn:f_imp}
f_{{\rm imp}}=\frac{\sum\left[ \left(\left|{f}_{\lambda,{\rm single}}-{f}_{\lambda}\right|-\left|{f}_{\lambda,{\rm binary}}-{f}_{\lambda}\right|\right)/{\sigma}_{\lambda}\right] }{\sum\left[ \left|{f}_{\lambda,{\rm single}}-{f}_{\lambda,{\rm binary}}\right|/{\sigma}_{\lambda}\right] },
\end{align}
where ${f}_{\lambda}$ and ${\sigma}_{\lambda}$ are the observed flux and corresponding uncertainty, ${f}_{\lambda,{\rm single}}$ and ${f}_{\lambda,{\rm binary}}$ are the best-fit single-star and binary model spectra, and the sum is over all wavelength pixels.% This improvement factor allowed us to select exposure where binary model has maximal performance

\subsubsection{Multi-epochs fitting}
\label{sec:multi}
We fit all available spectra for a given star, assuming that their spectral parameters are constant for all epochs.   
If the components of a binary system are gravitationally bound, their radial velocities should agree with the following equation \citep{wilson}:  

\begin{align}
\label{eqn:asgn}
    {\rm RV_2}=\gamma_{\rm dyn} (1+q_{\rm dyn}) - q_{\rm dyn} {\rm RV_1},
\end{align}
where $q_{\rm dyn}={M_1}/{M_2}$ - the mass ratio of the binary components and $\gamma_{\rm dyn}$ - the systemic velocity. Using this equation we can directly measure the systemic velocity and mass ratio. 
\par
To reduce the number of free parameters in multi-epochs fitting we use only ${\rm RV_1}$ and computed ${\rm RV_2}$ using Equation~\ref{eqn:asgn}. The same value of the mass ratio is used in Equation~\ref{eq:bolzmann}. Unlike the previous stage, we fit $\feh$ for both components. In total, we fit for 14 or 15 free parameters, for two and three epochs respectively.   
We use the previously derived $\rv$ to get initial guesses for $\gamma_{\rm dyn}$ and $q_{\rm dyn}$ using a linear fit. We fit all previously normalised individual epoch's spectra using their estimation of the spectral parameters for initialisation. We select the solution with minimal  $\chi^2$ as the final result.

\subsubsection{Typical errors estimation}
\label{sec:err}
We estimate typical errors of the multi-epochs fitting by testing it's performance on the dataset of synthetic binaries.  We generated 30 mock spectra for binary systems with the mass-ratio $q=1.5$, ${\teff}_{1,2}=12000,\,10000$ K, $\logg_{1,2}=3.9,\,4.0$ (cgs), ${\Vmic}_{1,2}=1\,\kms$ and $\vsini_{1,2}=10\,\kms$. Metallicity was set to $\feh=0.0$ dex for both components. Radial velocities were computed using circular orbit with the semiamplitude of the primary component $200\,\kms$ at randomly chosen phases. $\rv_2$ were computed using Eq.~\ref{eqn:asgn} with $\gamma=0~\kms$. Synthetic spectra were degraded by Gaussian noise according to $\snr=100$ pix$^{-1}$. 
\par
We performed exactly the same analysis as for the observations on this simulated dataset. We repeat multi-epochs fitting 100 times randomly choosing three spectra, previously analysed in individual spectra mode. We selected 88 good fits, where the error in fitted $\rv_1$ was less than $30\,\kms$ for all three epochs. We estimate how well parameters of the primary and secondary components can be recovered by calculating the average and standard deviation of the residuals. Among good solutions we have $\Delta \rv_1=0.02\pm0.15~\kms$ and $\Delta \gamma=0.16\pm1.06~\kms$. For $\teff$ we have $\Delta_{1,2}=13.26\pm19.79,\, 30.13\pm92.14$~K, for $\logg$ we have $\Delta_{1,2}=0.01\pm0.01,\, 0.03\pm0.03$ cgs units, for $\feh$ we have $\Delta_{1,2}=0.03\pm0.02,\, 0.08\pm0.09$ dex,  for $\Vmic$ we have $\Delta_{1,2}=0.24\pm0.11,\, 0.20\pm0.25$ $\kms$ and for $\vsini$ we have $\Delta_{1,2}=0.61\pm0.51,\, 2.62\pm13.84$ $\kms$. The recovered mass ratio has $\Delta q=-0.02\pm0.11$. It is clear that the parameters of the secondary component are less accurate compared to the parameters of the primary component and there are no significant biases in the recovered results.

\section{Results}
\label{results}%VV455Car and VANDor 
We fit all 67 individual epoch's spectra with the single-star and binary spectral models. We carefully inspect all spectral fits by eye and find that mostly the binary model is able to properly separate contributions of the two components in the composite spectrum. However for some spectra the binary model shows no significant improvement over the single-star model. These systems probably contain very similar components with very small $\rv$ separation.
For example the fits for two systems of fast rotating stars HD~29994 and AN~Dor are poor, although the binary model fits the spectra slightly better than the single-star model. Such fast rotators are very difficult targets for separation as all spectral lines are very broad. For a few spectra of $\epsilon$~Vol the binary model was completely ignoring the secondary component contribution, which was clearly visible, thus we measure $\rv$ by hand using the Mg~II line at 4481.33~\AA~and use these values to initialise the multi-epochs fitting.
\par%, and fast rotators shown in Figure
After visual inspection of the multi-epochs fitting results we select seven SB2 candidates listed in Table~\ref{tab:selected}.  We omit errors from the $\chi^2$ minimisation as they are largely underestimated. We also list $\rv$ measurements in Table~\ref{tab:all_specs}. Note that the $\rv$ derived in simultaneous fitting of multi-epochs can differ significantly from the values from individual spectra. Selected binaries can be splitted in two groups: slow rotators ($\vsini_{1,2}\leq70\,\kms$) with relatively narrow spectral lines shown in Figures~\ref{fig:spfit}, \ref{fig:spfit3} and binaries with discrepant $\vsini$ shown in Figure~\ref{fig:spfit2}. We zoom into regions with strong spectral features like the H$_\beta$ line core and the Mg~II line at 4481.33~\AA. Narrow spectral lines in stars with slow rotation show a significant RV shift between the two components. Our spectral model cannot fit individual chemical abundances, only $\feh$, therefore fit residuals are non zero (see for example region around Mg~II line at 4481.33~\AA). When a binary has components with very discrepant $\vsini$ (in Figure~\ref{fig:spfit2}), narrow lines of the slowly rotating component can be seen on top of the very broadened lines of the fast rotator. In such systems, the $\feh$ of the slow rotating component is more reliable, as the fast rotating component shows no good metal lines. For HD~208433 we have only two spectra taken at very similar phase, based on $\rv$, therefore the mass ratio and $\gamma$ for this system are less reliable than for the other SB2 candidates.
\par 

\begin{table*}
    \centering
    \begin{tabular}{lcccccccccccc}
\hline
id&$\gamma$&${\teff}_1$&${\teff}_2$&$\logg_1$&$\logg_2$&$\feh_1$&$\feh_2$&${\Vmic}_{1}$&${\Vmic}_2$&$\vsini_{1}$&$\vsini_{2}$&$q$\\
& $\kms$& K & K& dex & dex& dex & dex& $\kms$&$\kms$&$\kms$&$\kms$& \\ 
\hline
%HD 198174(*)& &13463&8706&3.2&3.02&\multicolumn{2}{c}{-0.31} &0.0&0.0&48.0&48.0&0.80\\
HD 20784&-20.79&9631&10292&3.43&3.80&0.01&-0.05&1.59&0.42&31&24&1.14\\
HD 208433&-4.64&11288&8528&4.23&4.31&-0.61&0.41&3.32&1.90&151&27&1.90\\
HD 43519&27.25&11145&9778&4.12&4.14&-0.1&-0.22&1.5&0.07&123&4&1.53\\
%HD 45796&5.20&15206&11319&4.05&3.93&0.02&-0.23&3.93&0.0&0.0&0.0&0.93\\
%HD 47620&30.23&17696&10364&4.2&3.62&-0.38&-0.19&11.03&0.0&0.0&26.0&1.79\\
HD 56024&7.22&12177&7954&3.81&3.74&-0.21&0.2&0.97&1.35&242&18&1.98\\
CD-73 375A&16.76&12181&9945&4.05&3.83&0.2&-0.25&0.43&0.21&7&23&1.53\\
AE Pic&29.27&17873&11699&3.93&4.02&-0.03&-0.28&0.0&1.13&70&42&1.43\\
$\epsilon$ Vol&6.94&14348&10772&3.62&4.07&-0.33&0.04&4.11&0.0&1&7&1.49\\ 
\hline

    \end{tabular}
    \caption{Spectral parameters for selected SB2 candidates.}% (*) based on single spectrum fit}
    \label{tab:selected}
\end{table*}

\begin{figure}
    \centering
    \includegraphics[width=\columnwidth]{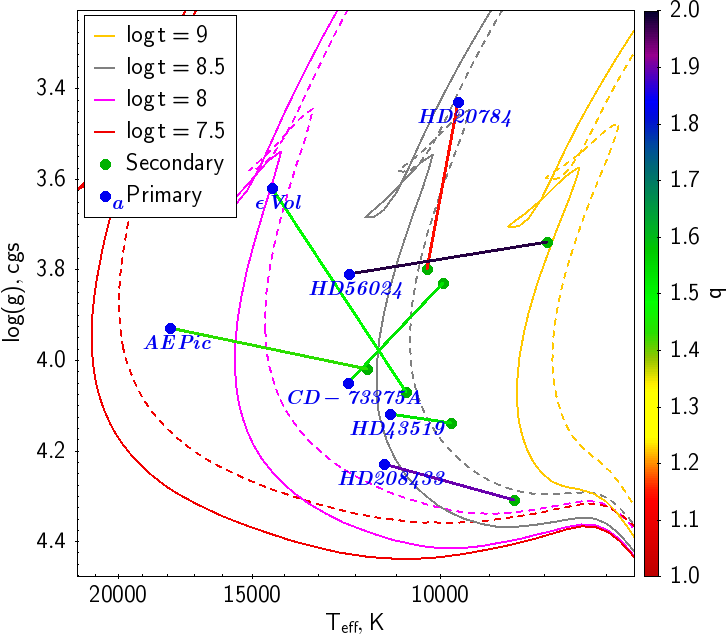}
    \caption{Kiel diagram for selected SB2 candidates. Primary and secondary components are connected by the line colored according to mass ratio. \texttt{PARSEC 1.2S} isochrones \protect\citep{marigo2017}, computed for four values of age, using $\feh=-0.3,\, 0.0$ dex are shown as solid and dashed lines, respectively.}
    \label{fig:kiel}
\end{figure}

\begin{figure}
    \centering
    \includegraphics[width=\columnwidth]{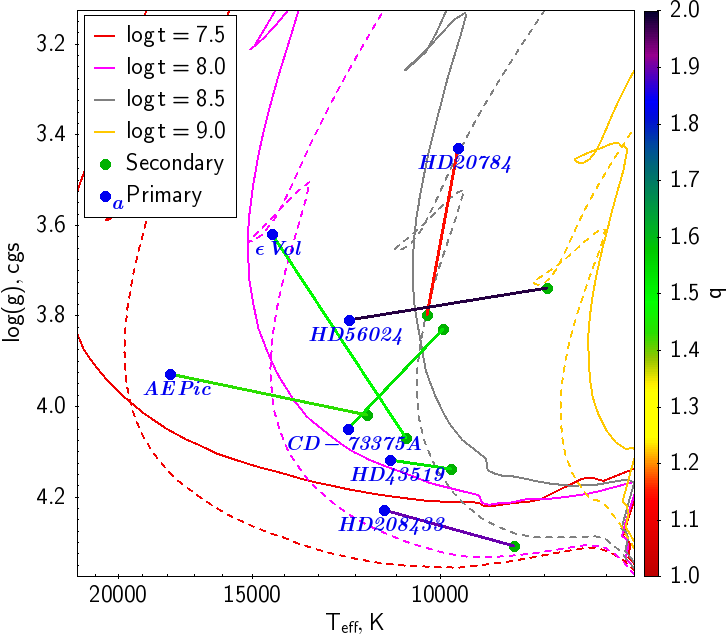}
    \caption{Kiel diagram for selected SB2 candidates. Primary and secondary components are connected by the line colored according to the mass ratio. Solar metallicity \texttt{PARSEC 2.0} isochrones \protect\citep{parsec2}, computed for four values of age, using rotation $\omega=0$ and $\omega=0.99\omega_{\rm crit}$ are shown as dashed and solid lines, respectively. Note that HD~20784 shows excellent agreement with the non-rotating isochrone of $\log{t}=8.5$ yr. }% $\epsilon$~Vol is
    \label{fig:kiel2}
\end{figure}

We compare the derived spectral parameters to \texttt{PARSEC~1.2S} \citep[][]{marigo2017} and \texttt{PARSEC~2.0} \citep{parsec2} stellar isochrones in Figures~\ref{fig:kiel}, \ref{fig:kiel2}. Overall agreement is good, better for \texttt{PARSEC~2.0}, six systems have a primary component that is hotter than the secondary, and only HD~20784 is an exception that seems to be more evolved. It shows excellent agreement with the non-rotating \texttt{PARSEC~2.0} isochrone of $\log{t}=8.5$ yr. This system will be discussed below in detail. $\epsilon$~Vol has significant $\Delta \logg \sim 0.5$ between the components, which can indicate that the primary has already crossed the turn-off point. HD~56024 shows a poor match to the values in the isochrones, having smaller $\logg$. This system also has the largest difference in mass and rotation between the components.

\subsection{HD 20784}
\label{hd20784}

\begin{figure}
    \centering
    \includegraphics[width=\columnwidth]{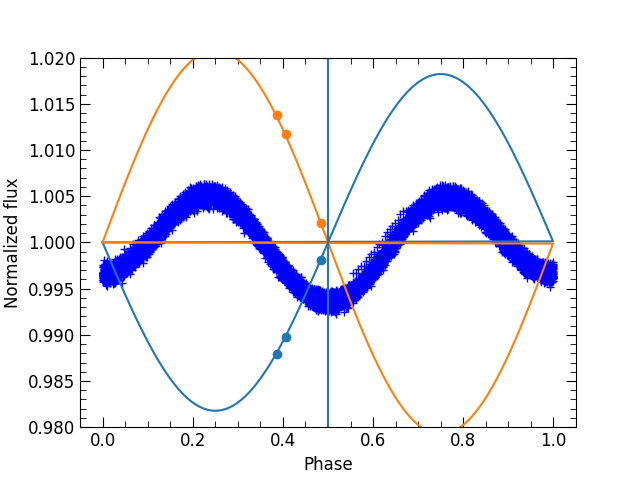}
    \caption{Folded TESS LC for HD~20784 with schematic plot of the circular orbit and radial velocities for the primary (from multi-epoch fit) and secondary (computed using Eq.~\protect\ref{eqn:asgn}) shown as blue and orange circles respectively. }
    \label{fig:lcfolded}
\end{figure}

\begin{figure}
    \centering
    \includegraphics[width=\columnwidth]{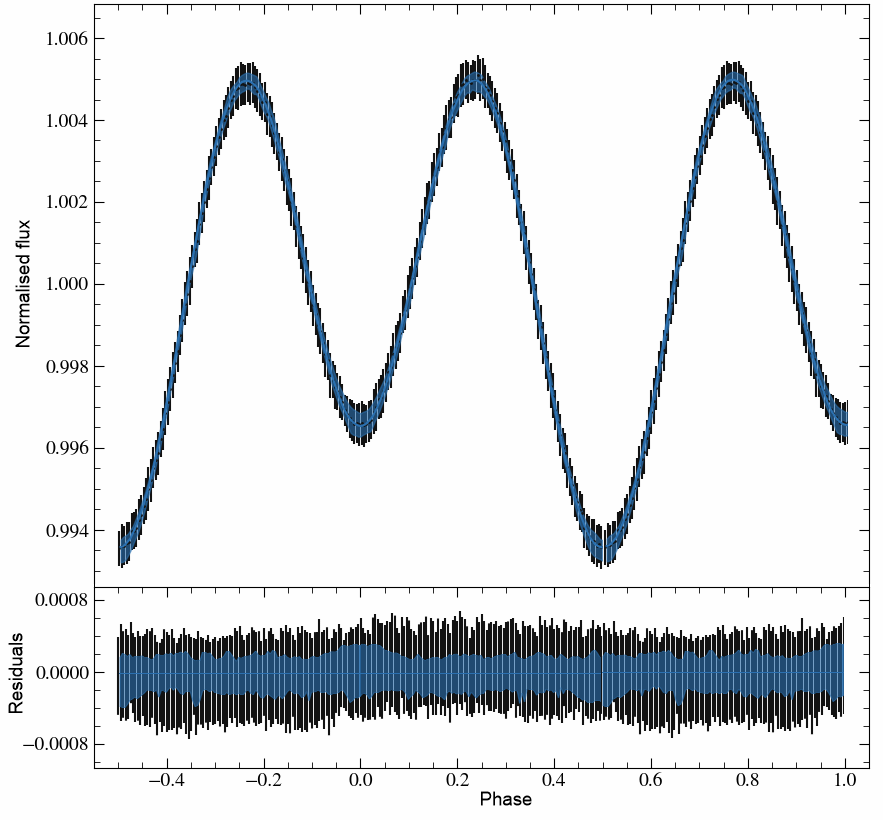}
    \caption{Best-fit \texttt{PHOEBE} model for the LC of HD~20784 with 500 sampled solutions, computed using $3\sigma$ uncertainties.}
    \label{fig:lcfit}
\end{figure}

We checked the MAST portal\footnote{\url{https://mast.stsci.edu/portal/Mashup/Clients/Mast/Portal.html}} and downloaded TESS light curves (LCs) from six sectors extracted by the Quick Look Pipeline (QLP) \citep[][]{tess1,tess2}. We used simple aperture photometry (SAP) with 11637 datapoints in total (selected by the quality flag) and find that the LC shows smooth sine-like variation with period $P\sim4$ d. \cite{2022ApJ...924..117B} measured rotation frequency $\nu=0.558\,d^{-1}$ for this system which gives us period $P=2/\nu=3.584$ d, which was used as an initial guess. To estimate the period and $t_0$ we selected the LC minima surrounded by at least 50 datapoints and measure their exact times $t_{min}$ by fitting parabolas. Then we adjust $P$ until $t_{min}$ folded with $P/2$ becomes constant versus time. Thus we find $P=3.5822389$ and $t_0=2457001.5528632$ d (see Figure~\ref{fig:period}). We show the folded LC and schematic circular orbit with three RV measurement in Figure~\ref{fig:lcfolded}. The RV and LC timeseries are consistent with synchronous rotation of a non-spherical primary component irradiated by a hotter secondary. We use binning to get 200 datapoints for the final LC, which was fitted by \texttt{PHOEBE} \citep{phoebe}. The resulting fit is shown in Figure~\ref{fig:lcfit}. We estimate uncertainties using 16 walkers and 1000 iterations in \texttt{emcee} sampling  \citep{emcee}. The final solution with $3~\sigma$ uncertainties is presented in Table~\ref{tab:phoebe}. Based on this, the primary fills a significant fraction of its Roche lobe (${R_{\rm equiv}}_1 \sim 0.75{R_{\rm equiv}}_1^{max}$), but mass transfer has not started yet. Ellipsoidal variability is not very sensible to the inclination (see \cite{tyc} with similar shape of LC), therefore the uncertainties are quite large. Taking them into account, the parameters of the components in HD~20784 agree well with spectroscopic estimates and with the values in \texttt{PARSEC} isochrones. Therefore our estimation of age is reliable.
%3.585

\begin{table}
    \centering
    \begin{tabular}{l|ccc}
\hline
Parameter & value& \texttt{PARSEC~2.0}& spectra\\
\hline
fixed:\\%LD 0.17 0.24 claret 2017
$P$, d& 3.5822389\\
$t_0$ (BJD), d& 2457001.5528632\\%t0 1.552863204576127 pm 0.14753752929325842
fitted:\\%K 63.76779189 gamma -0.5469111
$a\sin{i},\,R_\odot$ & $9.7^{+3.6}_{-3.2}$\\
$i^{\degr}$ & $26.7^{+7.3}_{-4.7 }$\\
$Q=1/q$ & $0.90^{ +0.66}_{-0.32}$ &0.93&0.88\\
${\teff}_1$, K & $9900^{+1100 }_{ -1500}$&9706&9631\\
${\teff}_2$, K & $10600^{+1600 }_{-1400 }$&10360&10292\\
${R_{\rm equiv}}_1,\,R_\odot$ & $6.03^{+1.38 }_{-0.66 }$&5.511\\%logL=4logT+2logR => R=10^(0.5*())
${R_{\rm equiv}}_2,\,R_\odot$ & $3.49^{ +0.97}_{-1.17 }$&3.52 \\
$L_3$, percent & $9.3^{+8.1 }_{ -6.9}$\\
derived:\\
$\logg_1$, cgs & $3.55^{+0.18 }_{-0.19 }$&3.456&3.43\\  
$\logg_2$, cgs & $4.14^{ +0.51}_{-0.60 }$&3.814&3.80 \\  
$M_1,\,M_\odot$ & $4.4^{ +5.7}_{ -1.9}$&3.163\\
$M_2,\,M_\odot$ & $3.5^{ +8.1}_{ -1.4}$&2.941\\
    \end{tabular}
    \caption{LC solution for HD~20784 with values from spectra and \texttt{PARSEC~2.0} isochrone, computed at solar metallicity, $\log{t}=8.5$ yr and $\omega=0$. }
    \label{tab:phoebe}
\end{table}

\section{Conclusions}
\label{concl}
% The last numbered section should briefly summarise what has been done, and describe
% the final conclusions which the authors draw from their work.

We successfully applied the composite spectral model to the high-resolution spectral sample of hot B stars observed by FEROS. Simultaneous fitting of multi-epoch spectra allows the method to get reliable estimates of the mass ratios and systemic velocities even from the two epochs, taken at phases with large $\Delta\rv$. We confirmed seven SB2 candidates out of 26 and found that our estimations of spectral parameters are consistent with \texttt{PARSEC} isochrones. We used TESS photometry of HD~20784 to verify our results and found good agreement between the LC modelling and spectroscopic analysis.
However we should note that the current version of the binary model is not optimal. A major part of the original sample was fitted poorly, despite the fact that all stars were preselected as SB2 by least-squares deconvolution analysis. Also the spectral model used only $\feh$ and cannot fit for abundances of individual chemical elements.  

\section*{Acknowledgements}

% We are grateful to the anonymous referee for a constructive report. We thank Hans B{\"a}hr for his careful proof-reading of the manuscript.
MK is grateful to his parents, Yuri Kovalev and Yulia Kovaleva, for their full support in making this research possible. The work is supported by the Natural Science Foundation of China (Nos. 11733008, 12090040, 12090043). 
The authors gratefully acknowledge the “PHOENIX Supercomputing Platform” jointly operated by the Binary Population Synthesis Group and the Stellar Astrophysics Group at Yunnan Observatories, Chinese Academy of Sciences. 
SG gratefully acknowledges support from the Research Foundation Flanders (FWO) by means of a PhD Aspirant mandate under contracts No.\,11E5620N.
This research has made use of NASA’s Astrophysics Data System, the SIMBAD data base, and the VizieR catalogue access tool, operated at CDS, Strasbourg, France. It also made use of TOPCAT, an interactive graphical viewer and editor for tabular data \citep[][]{topcat}.  Funding for the TESS mission is provided by NASA’s Science Mission directorate. This paper includes data collected by the TESS mission, which is publicly available from the Mikulski Archive for Space Telescopes (MAST).
The research leading to these results has (partially) received funding from the European Research Council (ERC) under the European Union’s Horizon 2020 research and innovation programme (grant agreement \textnumero 670519: MAMSIE) and from the KU Leuven Research Council (grant C16/18/005: PARADISE).

%%%%%%%%%%%%%%%%%%%%%%%%%%%%%%%%%%%%%%%%%%%%%%%%%%
\section*{Data Availability}
The data underlying this article will be shared on reasonable request to the corresponding author.

% The inclusion of a Data Availability Statement is a requirement for articles published in MNRAS. Data Availability Statements provide a standardised format for readers to understand the availability of data underlying the research results described in the article. The statement may refer to original data generated in the course of the study or to third-party data analysed in the article. The statement should describe and provide means of access, where possible, by linking to the data or providing the required accession numbers for the relevant databases or DOIs.



\bibliographystyle{mnras}

%\begin{thebibliography}{99}
%\end{thebibliography}

%%%%%%%%%%%%%%%%%%%%%%%%%%%%%%%%%%%%%%%%%%%%%%%%%%

%%%%%%%%%%%%%%%%% APPENDICES %%%%%%%%%%%%%%%%%%%%%

\appendix

\section{Some extra material}
\begin{figure*}
    \centering
    \includegraphics[width=\textwidth]{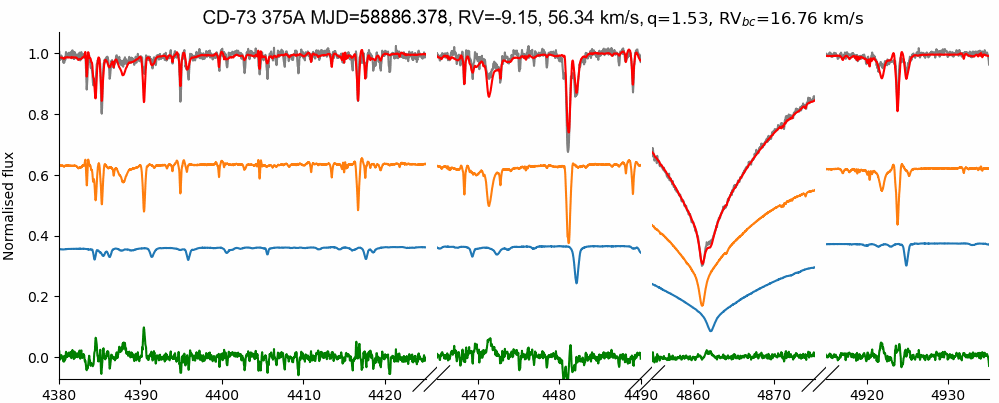}
    \includegraphics[width=\textwidth]{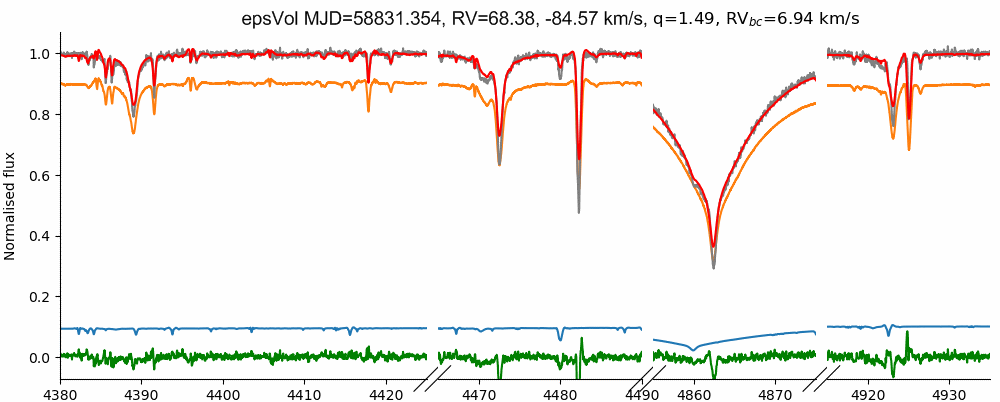}
    \includegraphics[width=\textwidth]{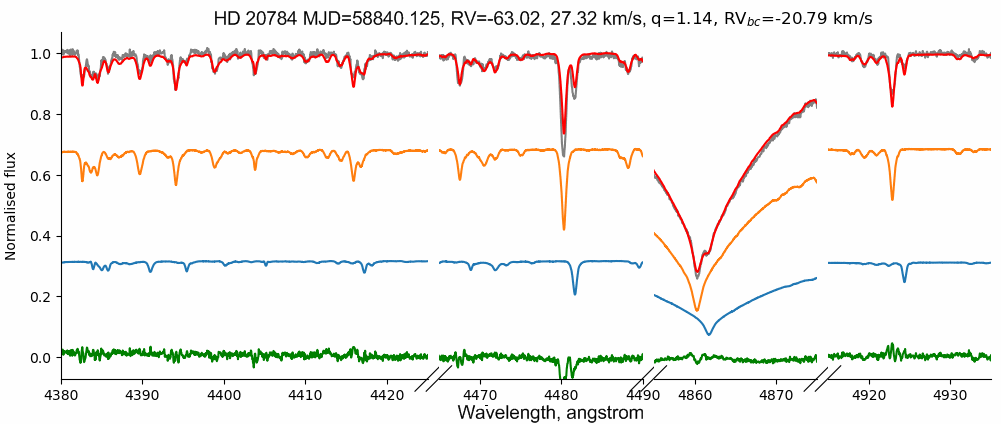}
    \caption{Example of the multi-epoch spectra fitting for three stars with slow rotation. The observed spectrum is shown as a gray line, the best fit is shown as a red line. The primary component is shown as the orange line, the secondary as a blue line. The difference O-C is shown as a green line.}
    \label{fig:spfit}
\end{figure*}

\begin{figure*}
    \centering
    \includegraphics[width=\textwidth]{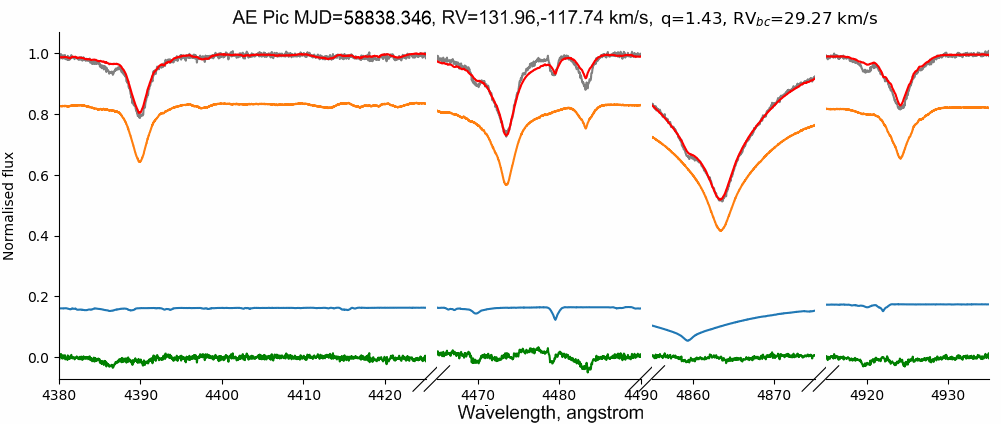}
    \caption{Same as Figure~\protect\ref{fig:spfit} for an SB2 system of two hot components}%fast rotating components. The observed spectrum is shown as a gray line, the best fit is shown as red line. The primary component is shown as the orange line, the secondary as a blue line. The difference O-C is shown as a green line.}
    \label{fig:spfit3}
\end{figure*}

\begin{figure*}
    \centering
    \includegraphics[width=\textwidth]{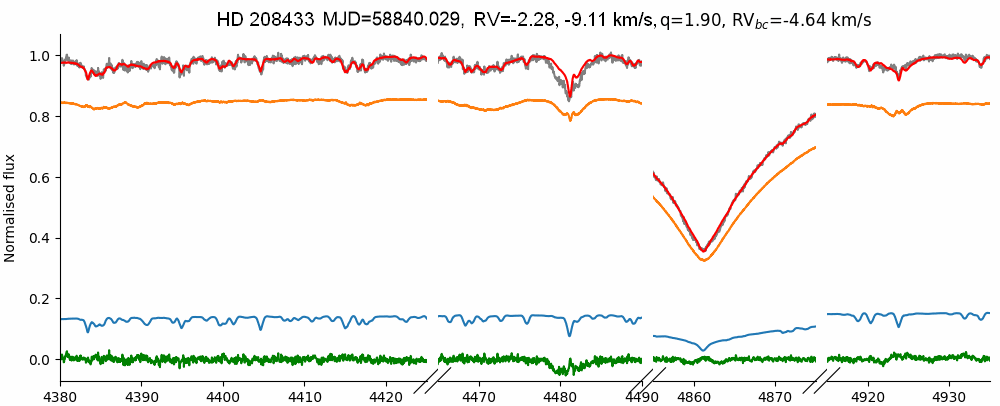}
    \includegraphics[width=\textwidth]{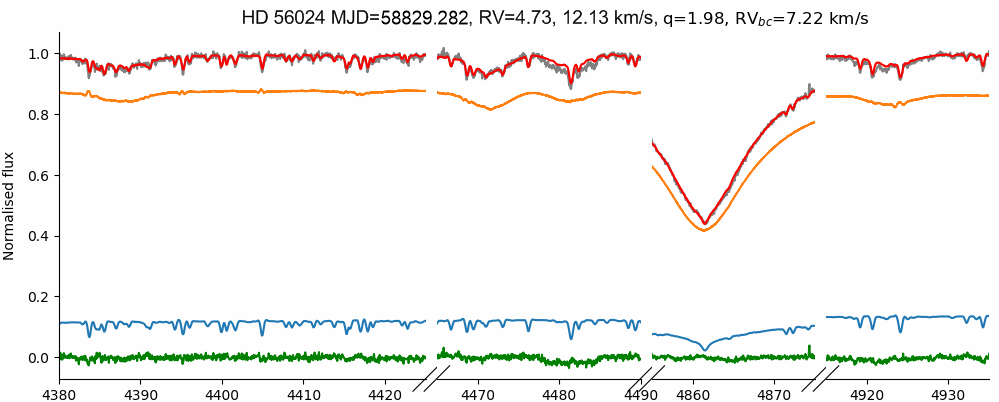}
    \includegraphics[width=\textwidth]{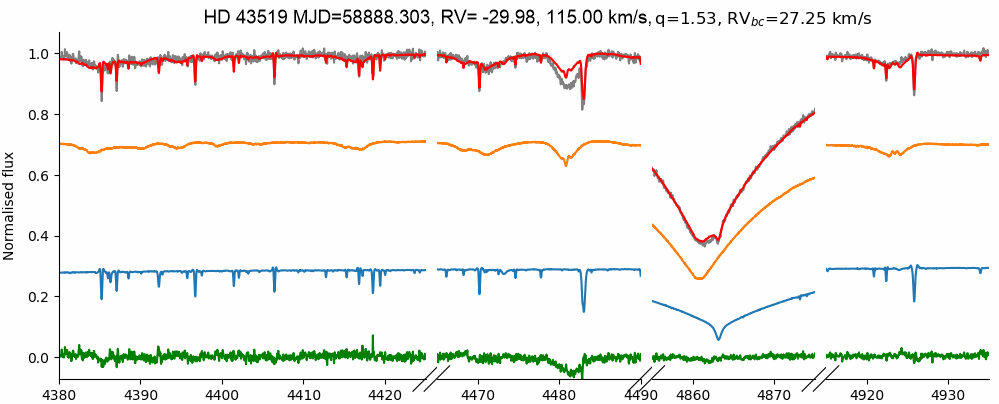}
    \caption{Same as Figure~\protect\ref{fig:spfit} for three stars with a fast rotating primary and a slow rotating secondary.}% The observed spectrum is shown as a gray line, the best fit is shown as red line. The primary component is shown as the orange line, the secondary as a blue line. The difference O-C is shown as a green line.}
    \label{fig:spfit2}
\end{figure*}

\begin{figure}
    \centering
    \includegraphics[width=\columnwidth]{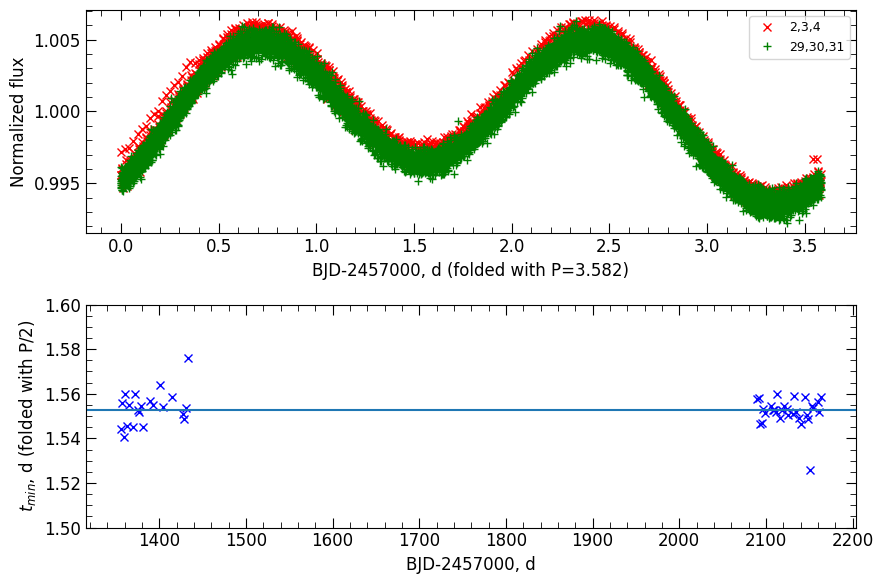}
    \caption{Folded TESS LC for HD~20784 from different sectors shown as green and red crosses respectively (top panel). $t_{min}$ folded with $P/2$ versus observation time are shown on lower panel. $t_0$ computed as median is shown as horizontal line.}
    \label{fig:period}
\end{figure}

Radial velocities for selected stars are listed in Table~\ref{tab:all_specs}. Figures~\ref{fig:spfit},\ref{fig:spfit3},\ref{fig:spfit2} show examples of best fits for selected SB2 candidates. Figure~\ref{fig:period} illustrates period finding procedure.

\begin{table*}
    \centering
    \begin{tabular}{l|ccccccccc}
    \hline% $\chi^2_{\rm single}$ &$\chi^2_{\rm binary}$
    id   & MJD & RV$^{\rm mult}_1$ & RV$^{\rm mult}_2$ & RV$_1$ & RV$_2$ & $f_{\rm imp}$ &$\vsini_{\rm single}$ & $\feh_{\rm single}$& frac\\
         &  d  & $\kms$            & $\kms$            & $\kms$ & $\kms$ &             &     $\kms$             &  dex &      \\
    \hline
HD 43519 &58837.336 &22.75 & 34.16 & 26.65 $\pm$ 0.09 &34.11 $\pm$ 0.05 &0.225 &83 &-0.09 &0.66\\
HD 43519 &58888.303 &-29.98 & 115.00 & -30.00 $\pm$ 0.08 &115.01 $\pm$ 0.04 &0.224 &167 &-0.10 &0.72\\
HD 20784 &58833.032 &-56.68 & 20.10 & -59.16 $\pm$ 0.06 &-8.94 $\pm$ 0.08 &0.332 &31 &0.04 &0.52\\
HD 20784 &58840.125 &-63.02 & 27.32 & -63.15 $\pm$ 0.04 &26.34 $\pm$ 0.10 &0.456 &67 &0.00 &0.72\\
HD 20784 &58887.049 &-27.27 & -13.41 & -33.46 $\pm$ 0.09 &-11.40 $\pm$ 0.08 &0.146 &30 &-0.00 &0.56\\
HD 208433 &58829.045 &-2.12 & -9.42 & -0.65 $\pm$ 0.07 &-10.06 $\pm$ 0.14 &0.291 &116 &0.02 &0.82\\
HD 208433 &58840.029 &-2.28 & -9.11 & -1.99 $\pm$ 0.05 &-9.75 $\pm$ 0.12 &0.343 &161 &-0.01 &0.83\\
HD 56024 &58829.282 &4.73 & 12.13  & 1.80 $\pm$ 0.05 &12.20 $\pm$ 0.06 &0.406 &209 &0.03 &0.86\\
HD 56024 &58885.344 &-27.07 & 74.95 & -83.37 $\pm$ 0.22 &76.01 $\pm$ 0.17 &0.103 &196 &-0.32 &0.60\\
CD-73 375A &58838.273 &43.58 & -24.20 & 43.37 $\pm$ 0.04 &-26.03 $\pm$ 0.10 &0.328 &69 &0.16 &0.71\\
CD-73 375A &58886.378 &-9.15 & 56.43  & -9.16 $\pm$ 0.03 &56.81 $\pm$ 0.10 &0.308 &81 &-0.03 &0.60\\
AE Pic &58838.346 &131.96 & -117.74 & 131.91 $\pm$ 0.05 &27.36 $\pm$ 0.13 &0.260 &103 &0.01 &0.65\\
AE Pic &58839.248 &37.04 & 18.15 & 34.91 $\pm$ 0.06 &24.88 $\pm$ 0.17 &0.097 &66 &-0.07 &0.70\\
AE Pic &58888.320 &13.70 & 51.57 & 16.18 $\pm$ 0.06 &38.79 $\pm$ 0.26 &0.099 &81 &-0.09 &0.85\\
$\epsilon$ Vol &58831.354 &68.38 & -84.57 & 68.60 $\pm$ 0.03 &1.79 $\pm$ 0.33 &0.126 &9 &-0.06 &0.85\\
$\epsilon$ Vol &58839.320 &-59.95 & 106.55 & -59.93 $\pm$ 0.01 &25.04 $\pm$ 0.10 &0.210 &9 &-0.09 &0.85\\
$\epsilon$ Vol &58883.379 &-48.13 & 88.95 & -48.14 $\pm$ 0.01 &15.41 $\pm$ 0.11 &0.159 &9 &-0.10 &0.85\\
\hline
    \end{tabular}
    \caption{Radial velocities for all spectra of selected SB2 candidates. frac=$k_{5000}/(1+k_{5000})$ is the primary light contribution at $\lambda=5000$~\AA \ from individual epoch fits.}
    \label{tab:all_specs}
\end{table*}

\subsection{ HD~198174}
\label{hd198}

\begin{figure}
    \centering
    \includegraphics[width=\columnwidth]{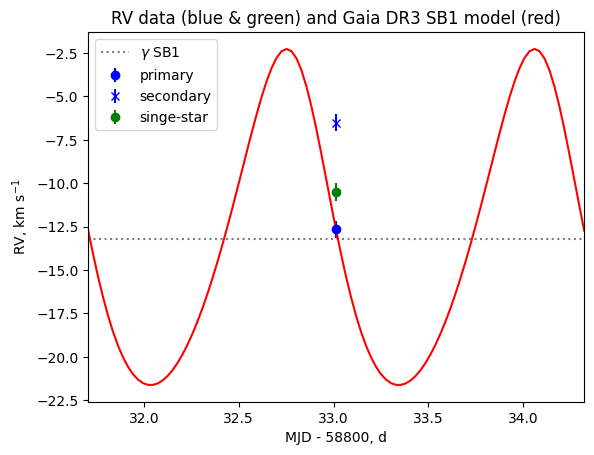}
    \caption{Gaia DR3 SB1 orbit and RV measurements from the FEROS spectrum of  HD~198174.}
    \label{fig:gaia3orbit}
\end{figure}

We cross match our set with the recent catalogue of non-single stars orbits from Gaia DR3 \citep[][]{gaia_dr3multiple} and find only one matching SB1 orbit for HD~198174 (period $P=1.3071\pm0.0001$ d, center of mass velocity $\gamma=-13.20\pm0.41~\kms$, periastron longitude $\omega=30.28\pm20.34$ deg, semiamplitude $K=9.68\pm0.53~\kms$, eccentricity $e=0.15\pm0.06$, periastron passage time (MJD) $t_0= 55196.48139\pm0.09287$ d). Our estimation of RV$_1=-12.66 \pm 0.10 \,\kms$ is in excellent agreement with this orbital solution, see Figure~\ref{fig:gaia3orbit}. However, the RV of the secondary component is not consistent with the SB1 orbit as it should be in opposite phase relative systemic velocity. Therefore our spectroscopic solution for this star is not reliable. %Using $\gamma$ and RV$_2=-3.58 \pm 0.16\,\kms$ we can derive mass ratio $q=16.8/2.4$%not clear how to  convert t0 to mjd 

\begin{figure*}
    \centering
    \includegraphics[width=\textwidth]{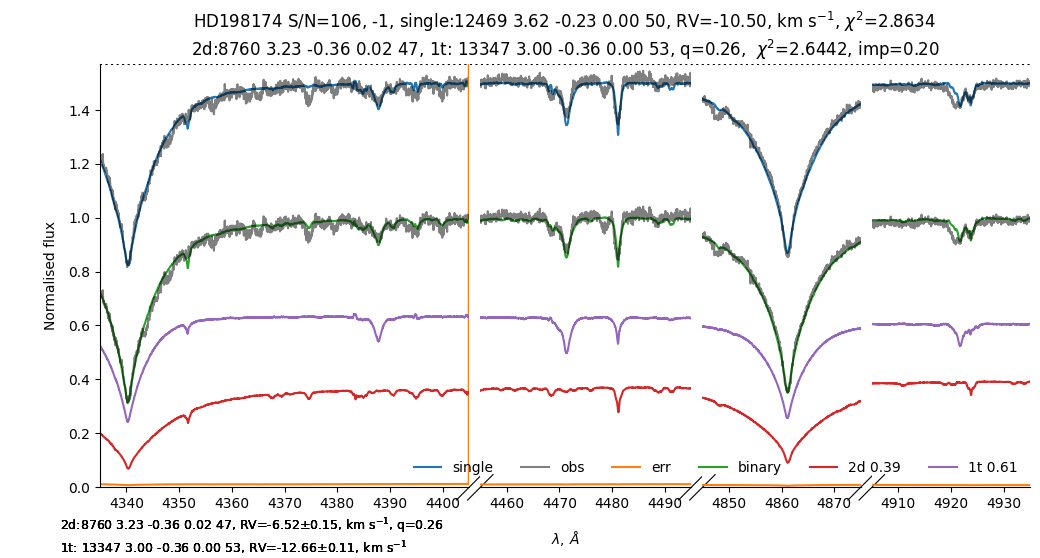}
    \caption{Example of the single spectrum of HD~198174 fitted by the binary spectral model (green line) and by the single star model (blue line) (with offset 0.5). The observed spectrum and its error are shown as gray and orange lines respectively. The primary (magenta line) and secondary (red line) components are labeled as "1t" and "2d" with their contribution to total light at $\lambda=5000$~\AA.~Spectral parameters ($\teff,~\logg,~\feh,~\Vmic,~\vsini$) from the single-star model fit and binary model fit are shown in the title.}
    \label{fig:single}
\end{figure*}

In Figure~\ref{fig:single} we can clearly see that the cool component ($\sim35$ percent of light) allows the binary model to fit the spectrum of HD~198174 better than the single-star model. Unfortunately it has only one spectrum, therefore the mass ratio is not reliable and this star is not selected as a good SB2 candidate.

%%%%%%%%%%%%%%%%%%%%%%%%%%%%%%%%%%%%%%%%%%%%%%%%%%

% Don't change these lines
\bsp	% typesetting comment
\label{lastpage}
\end{document}